# Proposal of an enriched three-tier test to assess learning risks in students on undergraduate physics courses

Ricardo Buzzo[1] and Alicia M. Montecinos[2]
[1] Institute of Physics, [2] MDCE Group,
Science Faculty, Pontificia Universidad Católica de Valparaíso, Avenida Universidad 330 Curauma, Valparaíso, Chile.

This proposal presents a methodology called *enriched three-tier test*, based on a similar test previously discussed in the literature. Ours consists of the use of *justification* and *degrees of confidence* combined with a multiple-choice test. This methodology of assessment allows the teacher to diagnose learning risks, while engaging the student in a metacognitive process. The three-tier test proved to be an efficient mechanism for identifying up to eighteen tints along the learning spectrum for a single concept, which interpretation alerts the teacher about the existence of learning risks and other learning situations. In this proposal, we explain the three-tier structure; we comment on it's application to undergraduate students from two semestral courses, and we analyze the category spectrum that a three-tier test produces, discussing it's differences with a traditional multiple choice question, it's limitations, and possible new applications.

## I. INTRODUCTION

Evaluation methods consisting of multiple-choice tests possess certain characteristics that make them an interesting option; these include the need for validation, the speed of qualification, the ease of statistical analysis of the results, and the possibility of evaluating a specific piece of content in an isolated and detailed manner. Mechanisms to evaluate the equity, validity and reliability of a multiple-choice test are widely described in the literature. However, their use by teaching staff is surprisingly uncommon, with the main reasons being a lack of knowledge and time [4]

Though the structure of this type of evaluation means that it is not possible to directly analyze any type of evidence developed by the student from a cognitive perspective, which leans the student towards a particular alternative and not the others, the distractors must be based on alternative models or conceptual errors reported in the literature, constituting plausible alternatives that are not easily discarded by the student [1, 4, 12]. Thus, the evaluator can identify which type of alternative model or conceptual error the student has applied. However, it does not allow identification of the *specific reasoning* undertaken by a particular student, irrespective of the alternative he/she chose (correct or incorrect).

Understanding this, some teachers consider it desirable to have more detailed information on the reasoning of students. A good example is in offering them the opportunity to justify their choice. A quantitative study was conducted by Dodd and Leal [6], concluding that the students feel comfortable with this methodology, as they receive some points for an incorrect answer, feeling that the teacher has not chosen the questions at random and that there is real interest from the teacher to find out what they have learned. However, the limitation of this study was that it was optional for the students to include their justification, and was only taken into account when they selected an incorrect answer.

If we now focus on the process of learning, the research of Scouller [18] confirmed that students have a greater propensity to use a superficial learning focus when they know they will be evaluated by traditional multiple-choice tests, seeing this type of evaluation as a test on the lowest levels of intellectual processing, i.e. remembering and reproducing. The researcher found a clear correlation between obtaining a good result and using superficial study strategies. The most alarming result of this study is that it correlated poor performance in a traditional multiple-choice test with the use of deeper learning strategies, where the latter is understood as the transformation of knowledge, reflection, and the use of arguments, among others.

Finally, we can add the well-known discussion on the effect of randomness on the result of a test of this type, as the students can guess or simply randomly choose their answer from only 4 or 5 alternatives. In order to decrease the influence of randomness, the basic recommendation is to extend the length of the exam (using a minimum of 60 questions) and using 5 alternatives and not 4, though statisticians state that the improvement is minor. Another method recommended along with the two above is to discourage guessing by taking points off for the number of wrong answers and rewarding the student when a question is left blank [2]. However, the difficulties implied by these safeguards are apparent.

Based on the above, the question that arises is how to improve this type of evaluation method while maintaining its benefits and avoiding its disadvantages. Also, turning it into a metacognition tool for the student and a meta-evaluation tool for the teacher, though these are not new questions (see, for example, [8, 10])

The present proposal describes an evaluation method that allows the teacher to ponder over the level of learning of the students in a more profound and efficient manner, which sheds light on the reasoning that is taking place. In particular, we concentrate on how to use an evaluation to identify the learning risks in university students on undergraduate physics courses.

The proposal comprises a test inspired by the three-tier test presented by Peşman and Eryilmaz [16] for the case of electric circuits which, in our case, will be applied to the subjects of mechanics and electromagnetism. The proposal is also based on the Degrees of Certainty Principle, widely studied by Dieudonné Leclercq (see, for example, Leclercq and Poumay [11])

We will explain how to describe the cognitive spectrum of students through 18 response types grouped into 7 categories, one of which can be used specifically to identify learning risks. The results and implications of the proposal are also discussed.

**II. THEORETICAL FRAMEWORK**

Leclercq and Poumay [op. cit.] operationally define metacognition as *judgment, analysis and/or observable regulations made by the student on their own performance (learning processes or products), either before, during or after, fundamentally in evaluations and learning*.

In the case of Degrees of Certainty (considered as a mode of metacognition by the aforementioned researchers), these are a *judgment* made *during* the evaluation, in which the student is explicit about his/her degree of certainty regarding their response. A case in which this mode is used is the work by Pérez de Landazábal, Benegas, Cabrera, Espejo, Macías, Otero, Seballos and Zavala [15] which quantified the initial level of knowledge of students that entered a total of 7 universities in Spain and Ibero-America. The research covered the topics of physics (mechanics and electricity) and mathematics (vectors, equations, derivatives) that are needed for the study of physics in introductory undergraduate courses. The exam that was used reported the degree of certainty on 4 levels: very certain, quite certain, some certainty and very little certainty, without greater detail on their usefulness or the projections for the method. The results of the study showed a completely unsatisfactory level on all areas evaluated transversally, despite the enormous differences in origin between the students in question.

Going further into the literature on learning problems in science, many examples can be found. Some specific examples from physics are Jones [9], Resnick, Chi, Slotta and Reiner [17], Eryilmaz [7], Campanario [3]. From them, we justify the *need* of the physics teacher to have a strategy for *detecting* the existence and persistence of learning risks among the students, alternative science models, naïve reasoning or conceptual errors.

This is fundamental, as there are many very effective methods for the theoretical/practical *teaching stage* in physics. One of the most successful methods is that developed by McDermontt, Mazur, Sokoloff and Thorton, who are considered pioneers in the development of methods whose effectiveness has been shown by well-known studies (see, for example, [5, 13,

19]). However, when looking for strategies for evaluation of the teaching-learning process or the *assessment stage*, there are fewer publications and the proposals are new and untested. One popular exception is the 4MAT methodology, which allows the teacher to *observe* how the students manifest their misconception, accordingly to their learning style [14].

Along the aforementioned line, there is the proposal by Peşman and Eryilmaz [16], who work with three-tier multiple-choice tests, combining the metacognitive mode of degrees of certainty with identification of learning risks. The structure of a question in a Peşman and Eryilmaz's three-tier test has the following three tiers:

1. *Question and alternative answers*. In the first part, as with a traditional test, the question is explained, including a graphical support if necessary, and the alternatives from which the student must choose are given.
2. *Identification of the reason*. In the second part the student is asked to select one of several justifications. One of them is the ideal justification but the rest are inconsistent with the acceptable science model. The option of writing out their own justification is also given.
3. *Identification of the degree of certainty*. In the third part the student can choose between two categories: Sure and Not Sure.

Unlike the above, the study of degrees of certainty by Leclercq originally used a scale with 6 levels: 0%, 20%, 40%, 60%, 80%, 100%, where 0% is an absence of certainty, and 100% is completely certain. The study by Pérez de Landazábal et al (op. cit.) also uses this mechanism. Due to the lack of a specific algorithm to allow students to quantify their degree of certainty, as there is no pattern for measuring certainty or confidence, we propose asking the student to report their certainty qualitatively, as explained in the methodology section below.

## III. METHODOLOGY

The enriched three-tier test was given to undergraduate students at the Pontificia Universidad Católica de Chile (PUCV Chile) on a semestral course on mechanics and one on electromagnetism; both part of different engineering degree programs. There were 4 evaluations for the mechanics course and 3 for the electromagnetism course, taking place during the second semester of the 2012 academic year.

**TABLE I**. Courses summary.

| Course | Number of Students | Number of Evaluations in which the enriched 3-tier test was used |
|---|---|---|
| Electromagnetism | 31 | 3 |
| Mechanics | 29 | 4 |

The structure of the questions in the enriched 3-tier test of our proposal is as follows:

1. *The question and alternatives.* Following the traditional model, this section includes the question, graphical support and the alternatives, where the distractors should follow the recommendations for a well-constructed question as set out in the introduction section above. The student is expected to select one alternative.

2. *Development.* A space is left for the student to write out a justification, either conceptual and/or mathematical, depending on the nature of the question.

3. *Degree of Certainty.* The student is expected to choose between **very certain**, **certain** and **uncertain**, to specify their degree of certainty and to initiate the act of metacognition.

The reason why the justification section (tier 2) does not include preset options in our proposal is in order to follow the essential idea of Dodd and Leal, allowing the teacher to add a partial score even when an incorrect answer has been selected. The method of Peşman and Eryilmaz would be very difficult for this action, as it would require selecting which alternative scientific model deserves a higher or lower score. In addition, this proposal does not aim to confirm the findings reported in the literature on errors in physics learning, but to value any partial knowledge on the part of the student.

In our proposal, the degree of certainty considers three levels. This decision is based on the aim of simplifying the structure of the question for the student, in order to efficiently involve the student in a metacognitive process which will end when said student receives their final evaluation results. Furthermore, as stated above, there is no way to measure degree of certainty numerically.

The following tables explain the structure of a three-tier question in our proposal (TABLE I and TABLE II).

**TABLE II.** Structure of an enriched three-tier question: description and explanation of how scores are assigned.

|  | *Question and alternatives* *0 or 1 point.* | *Development* *0 or 2 points.* | *Degree of Certainty* *No points assigned.* |
|---|---|---|---|
| Score assigned for qualification | 1 point is given for the correct answer. No negative points or deductions are given for incorrect answers, as this would be inconsistent with the second part and, as explained below, it is unnecessary. | 0, 1 or 2 points are given for the justification written out by the student, where 2 points are given for a perfect justification. Thus, each question has a maximum total of 3 points. | The students are told that no points are given for indicating their degree of certainty, but it is required. In other words, if this part is not completed, no points will be given for the first and second tiers. |

**TABLE III.** Structure of an enriched three-tier question: description and explanation of evaluation for learning risk analysis.

|  | *Question and alternatives* | *Development* | *Degree of Certainty* |
|---|---|---|---|
| Values for the metacognitive analysis and the detection of learning risks. | 0: Incorrect answer. 1: Correct answer. | 0: Justification absent or completely incorrect. 1: Justification partially correct. 2: Justification consistent with the ideal science model. | 0: Uncertain. 1: Certain. 2: Very certain. |

The technique of placing values for the three tiers described above gives a total of 18 response types. Each one is given a code or *key*, as described in the fourth column of table IV:

TABLE IV: Keys and description of the 18 categories.

| Tier 1: Question and alternative answers | Tier 2: Development | Tier 3: Degree of Certainty | Key for this Category | Description of the category |
|---|---|---|---|---|
| 0 | 0 | 2 | -9 | High degree of certainty in a wrong answer, showing the installation of an alternative model or conceptual error by the student. |
| 0 | 0 | 1 | -8 | Medium degree of certainty in a wrong answer, showing the installation of an alternative model or conceptual error by the student. |
| 0 | 0 | 0 | -7 | This is a neutral category. It refers to two possibilities: the student has opted not to answer the question, or is uncertain of the development, which is wrong. |
| 0 | 1 | 2 | -6 | High degree of certainty in a wrong answer, but with a development that partially responds to the ideal model. Though there is a certain level of risk, it cannot be said that there is installation of a conceptual error, but rather that different science models and alternatives are coexisting, or there may be errors in the application of the correct model, which would indicate an incomplete development of a certain level of ability. |
| 0 | 1 | 1 | -5 | Medium degree of certainty in a wrong answer, but with a development that partially responds to the ideal model. Though there is some risk, it cannot be said that there is installation of a conceptual error, but rather different science models or alternatives are coexisting, or errors in the application of the correct model, which would indicate the incomplete development of a certain level of ability. |
| 0 | 1 | 0 | -4 | This type of wrong answer shows uncertainty, showing a very weak level of useful knowledge. |
| 0 | 2 | 0 | -3 | False Negative with a low level of certainty. This type of wrong answer shows uncertainty, though paradoxically showing a good level of knowledge. It is interesting, as the reason why the student did not get the correct answer despite having the correct model should be explored. |
| 0 | 2 | 1 | -2 | False Negative with a medium degree of certainty. The student has the necessary knowledge, but is unable to identify the correct answer. |
| 0 | 2 | 2 | -1 | False Negative with a high degree of certainty. |
| 1 | 0 | 0 | 1 | A very interesting case in which the student is very certain of the development, but did not transfer the correct reasoning into identifying the correct alternative. |
| | | | | False Positive with a low level of certainty. It represents the selection of the correct answer either with the justification left blank or completely wrong, with a degree of certainty that shows that the alternative was chosen at random or by guessing. |
| 1 | 0 | 1 | 2 | False Positive with a medium level of certainty, which may indicate the selection of an alternative at random or by guessing. |
| 1 | 0 | 2 | 3 | False Positive with a high degree of certainty, which may indicate the selection of an alternative by guessing or because the student is applying memory. |
| 1 | 1 | 0 | 4 | This type of answer shows weak knowledge combined with uncertainty. |
| 1 | 1 | 1 | 5 | This type of answer shows weak knowledge combined with a medium degree of certainty. |
| 1 | 1 | 2 | 6 | This type of answer shows weak knowledge combined with a high degree of certainty. This is the typical case in which the student does not completely justify their response, or does so in an inadequate way, and cannot therefore be assigned a perfect value. |
| 1 | 2 | 0 | 7 | Perfect answer marked by a student who is uncertain. The lack of certainty needs to be analyzed. |
| 1 | 2 | 1 | 8 | Perfect answer, but with a medium level of certainty, showing that work needs to be done on the reason for the uncertainty. This could be modesty or that the student still has some doubts. |
| 1 | 2 | 2 | 9 | This category represents the ideal answer, in which the correct alternative is chosen, through a coherent argument, with a high level of certainty. |

Based on the analysis of the categories above, it is possible to separate them into 7 response groups, within which the zone of learning risk can be identified, as explained in table V:

**TABLE V.** Description of the seven different groups of responses.

| Group | Categories | Description |
|---|---|---|
| Good Certain Learning | 8 and 9 | The type of response that shows satisfactory learning. |
| Good Uncertain Learning | 7 | Show that though the student has the necessary tools to satisfactorily develop the question, it is not seen by the |
| Weak Learning | 4, 5 and 6 | student themselves, leading to uncertainty. This group of correct answers does not have a completely satisfactory argumentation, and has different levels of certainty. It is evidence that there are gaps, or even a conceptual error involved, which needs to be worked on. |
| False Positives | 1, 2 and 3 | This zone shows the type of correct answer that is chosen at random, by guessing or by memorizing a previous similar question, as there is a low score in the justification. |
| False Negatives | -1, -2 and -3 | This group includes the answers that were close to perfect, as they have the correct arguments, but problems in identifying the correct answer, as well as different degrees of certainty. It represents false negatives that show satisfactory knowledge. |
| Zone of Learning Risk | -5, -6, -8, and -9 | This group includes the answers that represent the probable presence of conceptual errors, alternative science models and weakly developed abilities, combined with medium and high degrees of certainty. **It represents the area of learning risk that the teacher must work with.** |
| Non-developed Learning | -4 and -7 | These are the two categories that represent answers with a low degree of certainty and show a lack of knowledge. It cannot be said that they include conceptual errors, as there is no certainty in the response. |

## IV. RESULTS

The following (FIG 1) is an example taken from the first assessment on the Mechanics Course, which evaluates whether the student can interpret the gradient of a line in a graph of position as a function of time as the speed for one-dimensional movement with constant acceleration, a very well known question for the teachers:

**Item 1 of the assessment 1, Mechanics Course.** This graph shows position as a function of time of two bodies A and B which move in one dimension. According to the graph, at which point are they moving at the same speed?

(A) Between 0 and 2 seconds.
(B) At 2 seconds.
(C) From 2 seconds onwards.
(D) Body A has the same speed throughout its movement as body B has in the first 2 seconds.
(E) Never.

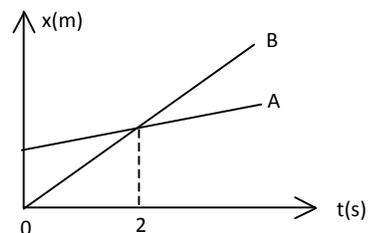

**FIG 1**. Example taken from the Mechanics Course's first assessment.

The results for this question are summarized in the following bar chart (FIG 2):

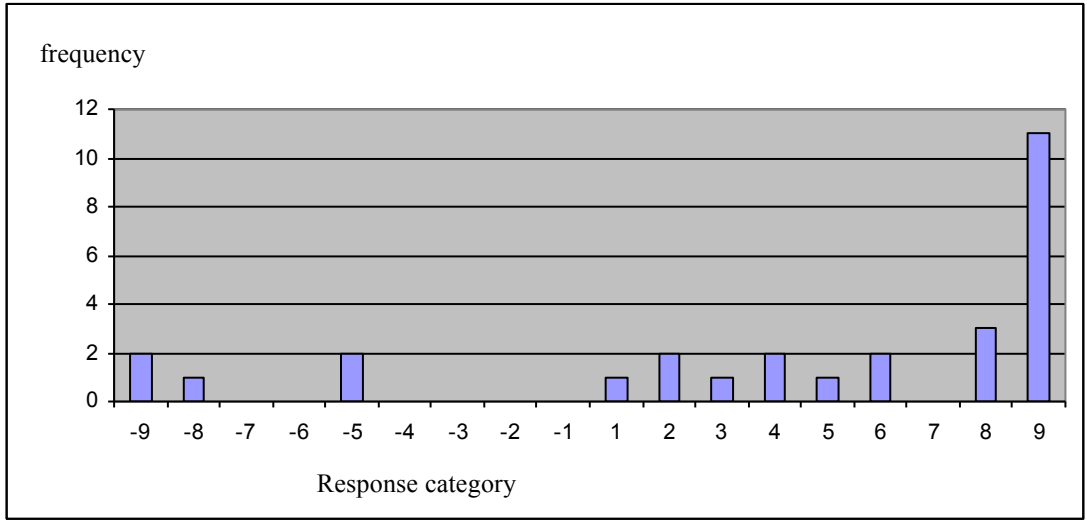

**FIG 2**. Results for item 1, assessment 1, Mechanics Course.

In this example it can be concluded that most of the class tends towards the ideal response. **75% of the class chose the correct answer**, and there were no omissions. Under traditional analysis of a multiple-choice test, we would conclude that most of the class can use the underlying concept of the gradient of a graph. However, the spectrum showed by our proposal can be used to conclude that **only 39% of the class had the ideal response** (category 9).

**In other words, only 48% of the students with the right answer also presented acceptable arguments for the model under evaluation**. Two students gave answers in the -9 category, and therefore were within the learning risk zone. The rest of the spectrum is mainly distributed over categories that show weak learning and even false positives, which shows the usefulness of not giving full points to an answer of this type.

It can be seen that the result from the traditional perspective is different that the result from our proposal, as are the implications for the work needed by the teacher, since our proposal reveals that educational decisions must be taken with regard to the physics concept under assessment.

Ideally, the results obtained in a complete test will be entered into a software package (spreadsheet) to perform a statistical analysis (for example, see the summary table in Crisp and Palmer [4] p 92). Each teacher can place emphasis on the statistical analysis of the category or set of categories as deemed necessary, either based on the questions, the subject matter, the course level, the individual level or per group of students.

Consider the following example provided in FIG. 3:

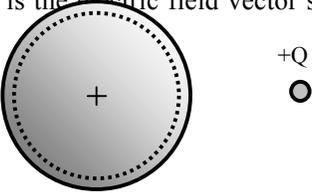

**Item 11 of the assessment 1, Electromagnetism Course**. The figure shows a hollow metal conductor sphere that initially has a positive charge distributed uniformly over its surface. If a particle with positive charge +Q is then placed close to the sphere, how is the electric field vector shown in the center of the sphere?
- A) it is directed to the left.
- B) it is directed to the right.
- C) it is directed up.
- D) it is directed down.
- E) it is zero.

**FIG 3**. Example taken from the Electromagnetism Course's first assessment.

The results of this question are summarized in the following bar chart (FIG 4):

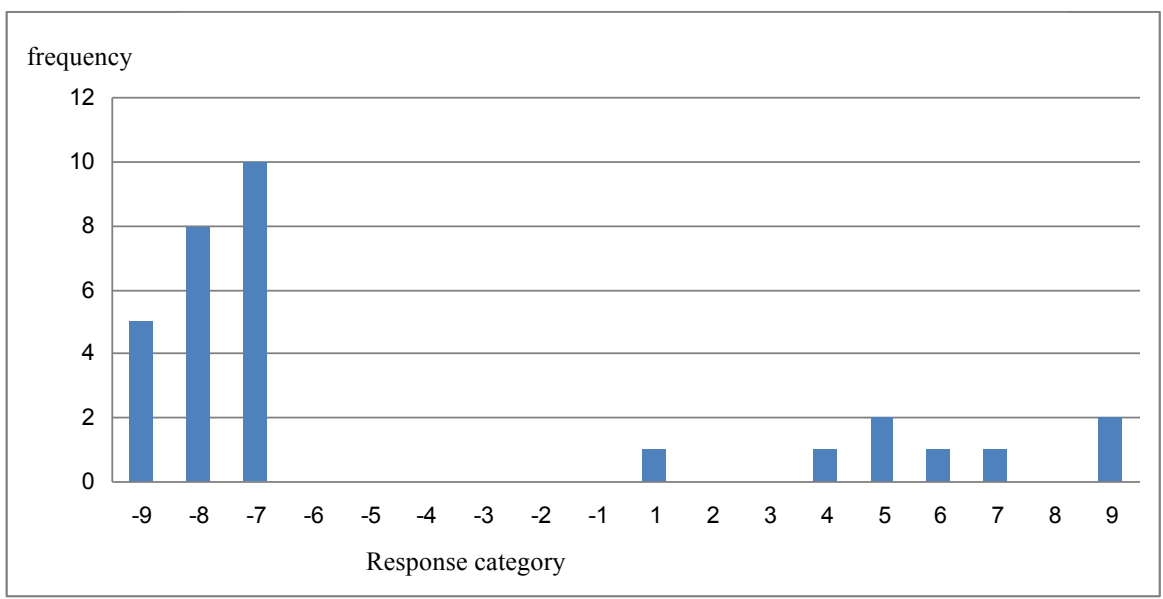

**FIG 4**. Results for question 11, assessment 1, Electromagnetism Course.

In this second case, it can be seen that 26% of the class chose the correct answer, and approximately 6% of the class gave the ideal response. There were 10 omissions (32% of the class) and it can be seen that all of the wrong answers fall in the learning risk zone, representing 42% of the class. Therefore, the teacher can differentiate the students that need to *learn* this subject (who may be those who omitted the question) from those who need a *conceptual change*, as they gave a medium and high degree of certainty in their wrong answers.

A. Didactic dimension emerging from the experience.

This methodology requires that the test papers are returned to the students, either temporarily or permanently, to allow the student to identify (using the instructions given by the teacher) the topics where they exhibit learning risks, thus **involving the student metacognitively**. The student does not need to be given the statistical indices, or the bar chart shown above. The students can freely compare with their classmates and consult the teacher as they deem necessary. The teacher must stimulate and guide the students in a reflection on what it was that led them into the zone of learning risk on certain topics, and **must guide them into a stage aimed at the correct conceptual change**.

**Finally, after an adequate amount of time the student must be given a new evaluation**, either accumulative or formative, in order to monitor the progress of the risks identified, where the strategy of this proposal can be repeated.

B. Analysis of False Positives.

As mentioned above, it is recommended to decrease the influence of the random effect in evaluations through a series of methods. In our study, the frequency of False Positives (categories 1, 2 and 3) is no more than 1% of the total number of responses. It is posited that this surprising statistic is the fruit of the very structure of the enriched three-tier test, which includes justification as an essential part of the evaluation process, discouraging the students from answering at random or from not justifying their response.

V. CONCLUSIONS AND DISCUSSION

From the analysis of the results of the evaluations, it can be said that this evaluation methodology is highly useful as it not only maintains the positive characteristics of a traditional multiple-choice test, but also adds a greater degree of analysis to what is found in the literature, i.e.:

- It allows the teacher to identify the spectrum of the class for a specific content, with a high level of detail, showing the learning risks and differentiating them from other types of phenomena. It particularly allows

- identification of a group of answers that indicate learning risks, differentiating them from a failure of learning, requiring different remedial action.

- It allows the student to develop metacognition, as it gives the possibility of reviewing how sure they are of their response to a question, what type of justification they used, and what was the ideal response, as well as other aspects. Involving the student in a dialogue with their peers is a sign of an active learning method, which implies that this proposal is coherent with the current lines in didactics.

The proposed three-tier test can enrich the academic relationship between students, and between the student and the teacher, since after analyzing the results of the evaluation through the spectrum obtained, the teacher can aim their specific actions at reverting the cases of learning risks, thus justifying to the students the reasons behind a new teaching direction.

The low percentage of False Positives supports the usefulness of the proposed method. It can be seen that using this strategy in the context of a multiple-choice evaluation can decrease the amount of questions needed, maintaining the validity of the instrument. Thus, it facilitates the task of the evaluator in terms of the time needed for the analysis, i.e. in correcting the tests.

It is posited that it may also be applied to the study of true/false questions.

Traditionally, when a question has a high rate of positive or negative answers, it was said that it was of low (or high) difficulty. Using a graphical frequency analysis per category is it possible to see finer *nuances* based on the degree of certainty and the quality of the arguments. In the case of unsatisfactory responses, it is possible to identify whether it is because the question was not understood correctly by the students, or perhaps because there was an event during the teaching-learning process.

Thus, it contributes to the meta-evaluative process of the teacher, who is able to efficiently discard, modify and improve questions for future applications.

It is clear that this type of evaluation methodology should be part of a bigger evaluation structure within the context of an academic semester.

One weakness of the methodology is in the case of analyzing the -7 category, the neutral category. A method for differentiating between omitted answers and answers with incorrect justification need to be designed, as these belong to very different processes.

**IV. ACKNOWLEDGMENTS**

The authors kindly regards the Physics Institute of PUCV Chile.